\documentstyle[prd,twocolumn,aps]{revtex}
\begin{document}
\draft
\title{The Free Quon Gas Suffers Gibbs' Paradox}
\author{R.F. Werner
\thanks{Electronic mail:
        \tt reinwer@dosuni1.rz.Uni-Osnabrueck.DE}
}
\address{FB Physik, Universit\"at Osnabr\"uck,
         Postfach 4469, D-4500 Osnabr\"uck, Germany.}

\date{April 20, 1993}

\maketitle

\begin{abstract}
We consider the Statistical Mechanics of systems of particles
satisfying the $q$-commutation relations recently proposed by
Greenberg and others. We show that although the commutation
relations approach Bose (resp.\ Fermi) relations for $q\to1$ (resp.\
$q\to-1$), the partition functions of free gases are independent of
$q$ in the range $-1<q<1$. The partition functions exhibit Gibbs'
Paradox in the same way as a classical gas without a correction
factor $1/N!$ for the statistical weight of the $N$-particle phase
space, i.e.\ the Statistical Mechanics does not describe a material
for which entropy, free energy, and particle number are extensive
thermodynamical quantities.
\end{abstract}


\section{Introduction }

In series of papers \cite{Greena,Green,Moha} O.W.~Greenberg and
collaborators have suggested a new way of interpolating between Bose
and Fermi statistics, which would presumably allow small deviations
from the standard description of identical particles. Their proposal
is to consider ``Quon'' creation and annihilation operators
satisfying the relations
\begin{equation} \label{e1}
  a(f)a^\dagger(g)-q\,a^\dagger(g)a(f)=\langle f,g\rangle\, \openone
\quad,
\end{equation}
where $f,g$ are test functions, i.e., elements of the one-particle
space with inner product $\langle f,g\rangle$, $a(f)$ is a Hilbert
space operator with adjoint $a^\dagger(f)$ depending linearly on
$f$. The deformation parameter $q$ must be real, and we take it in
the interval $-1<q<1$. At the endpoints of this interval the
relations become the canonical commutation relations ($q=1$), and
the canonical anticommutation relations ($q=-1$), respectively. The
Quons with $q=0$ were the first example of this type considered by
Greenberg \cite{Greena}. The relations (\ref{e1}) with $q=0$ also
play a central r\^ole in the theory of freely independent random
variables \cite{Maassen,Speia,Speib} initiated by Voiculescu
\cite{Voica}. Such systems have been used as a driving noise
\cite{Speia} in quantum stochastic differential equations
\cite{KuSpei}. The extension of this work to other values of $q$ has
led to an independent proposal of the relations (\ref{e1})
\cite{BoSpei}. The relations also arise as the commutation relations
of collective degrees of freedom in a system of many components
obeying anomalous statistics \cite{SpeMac}.

The Quon system with a single degree of freedom, for which the test
function space is  one-dimensional, and only one relation
$aa^\dagger-qa^\dagger a=\openone$ remains, is called a
$q$-oscillator. It naturally arises in the theory of quantum SU$_2$
\cite{Worono} and has also been introduced in this context by
Biedenharn \cite{Bieden} and MacFarlane \cite{Farlane}. The theory
of this system can be made to look very much like the theory of the
ordinary oscillator by means of ``$q$-analysis''
\cite{Fivel,Florea,Koorn}. A ``gas'' of $q$-oscillators can be
defined and studied using standard quantum mechanical procedures, as
soon as one decides whether these systems should be Bosons or
Fermions. The $q$-commutation relations then only enter as a way of
defining a special one-particle Hamiltonian (see e.g.\ \cite{Lee}).

In this note we consider the rather less trivial ``Quon gas''
according to Greenberg's original proposal, i.e.\ we demand that
different quonic degrees of freedom, corresponding to orthogonal
test functions $f,g$, also satisfy the relations (\ref{e1}). In
particular, we address the question of the $q$-dependence of the
partition function of a free Quon gas. We begin by establishing the
construction of Quon second quantization. We then derive the
formulas for computing partition functions and expectation values of
second quantized observables, and show that these formulas do not
contain the parameter $q$. Finally, we show that the Statistical
Mechanics of non-relativistic Quons exhibits the Gibbs' Paradox
known from Classical Statistical Mechanics. The main intention of
this paper is to state this Paradox as simply as possible, observing
accepted standard procedures of theoretical physics. However, at the
end we discuss some possible variations of the approach described in
the paper.

\section{ Quon Second Quantization}

The free Quon gas is a system described in a Fock Hilbert space
generated from a vacuum vector $\Omega_q$ with $a(f)\Omega_q=0$ by
successive application of creation operators $a^\dagger(f)$. The
scalar product of the vectors
$a^\dagger(f_1)\cdots a^\dagger(f_n)\Omega_q$,
which by definition generate the $q$-Fock space, can be computed
using only the commutation relations (\ref{e1}) and the condition
$a(f)\Omega_q=0$, and gives \cite{BoSpei}
\begin{eqnarray} \label{e2}
  \langle &&a^\dagger(f_1)\cdots a^\dagger(f_n)\Omega_q,\
            a^\dagger(g_1)\cdots a^\dagger(g_m)\Omega_q\rangle
\nonumber\\
\nonumber\\
          &&=\cases{0&if $n\neq m$\cr
               \displaystyle
               \sum_{\pi\in S_n} q^{I(\pi)}
                \langle f_1,g_{\pi(1)}\rangle\cdots\langle
f_n,g_{\pi(n)}\rangle
             & otherwise.}
\end{eqnarray}
Here the sum is over all permutations $\pi$ of $n$ elements, and
$I(\pi)$ denotes the number of inversions of the permutation $\pi$,
i.e. the number of pairs $(i,j)$ such that $i<j$ and $\pi(i)>\pi(j)$.
The positivity of the scalar product follows from the positive
definiteness of $q^{I(\pi)}$ as a function of $\pi$, shown in
\cite{BoSpei} (compare \cite{Fivel,Zagier}).
Completing with respect to this scalar product we obtain the Hilbert
space of a second quantized Quon system. For $q=0$ this space is
exactly the ``full Fock space'', i.e.\ the direct sum of the
unsymmetrized $n$-fold tensor products of the one-particle space. In
equation (\ref{e2}) we have followed the notational convention that
the $q$-dependence is attached to the vacuum vector $\Omega_q$, and
the creation and annihilation operators are denoted by the same
symbols for all $q$. The connection between $q$-Fock space and the
full Fock space at $q=0$ is then given by an operator $S_q$ with
\begin{equation} \label{e3}
 S_q\ a^\dagger(f_1)\cdots a^\dagger(f_n)\Omega_q
   =a^\dagger(f_1)\cdots a^\dagger(f_n)\Omega_0
\quad.
\end{equation}
It was shown in \cite{BoSpei,Zagier,Dykema} that, restricted to
the $n$-particle space, this operator is boundedly invertible for
all $-1<q<1$.

What makes a Quon system ``free'' is that its Hamiltonian and time
evolution are canonically given in terms of the corresponding
objects on the one-particle Hilbert space:
if $u_tf=e^{-it{\bf h}}f$ describes the time evolution of a single
Quon, the free time evolution on Fock space is given by
\begin{displaymath}
 U_t=e^{-iHt}
\end{displaymath}
\begin{equation} \label{e4}
\text{with}\qquad
   U_ta^\dagger(f)U_t^*=a^\dagger(u_tf)
\end{equation}
\begin{displaymath}
\text{and}\qquad
   U_t\Omega_q=\Omega_q
\quad.
\end{displaymath}
We will use the notations $U_t=\Gamma_q(u_t)$ and
$H={\rm d}\Gamma_q({\bf h})$ for this ``functor'' of second
quantization. In particular, ${\rm d}\Gamma_q(\openone)=N$ is the
number operator in Fock space.

We will now collect a few properties of $\Gamma_q$, which will be
useful later on. These hold for all $q$ including $q=\pm1$. We first
extend the definition of $\Gamma_q$ and ${\rm d}\Gamma_q$ from
unitary and self-adjoint operators to more general operators on
one-particle space by setting
\begin{mathletters}
\begin{equation} \label{e5a}
 \Gamma_q(R)a^\dagger(f_1)\cdots a^\dagger(f_n)\Omega_q
      =a^\dagger(Rf_1)\cdots a^\dagger(Rf_n)\Omega_q
\end{equation}
\begin{eqnarray} \label{e5b}
{\rm d}\Gamma_q({\bf h})&&a^\dagger(f_1)\cdots
                          a^\dagger(f_n)\Omega_q
\nonumber\\&&=
          a^\dagger({\bf h} f_1)a^\dagger(f_2)
          \cdots a^\dagger(f_n)\Omega_q
\nonumber\\&& \quad\qquad +
          a^\dagger(f_1)a^\dagger({\bf h} f_2)
         \cdots a^\dagger(f_n)\Omega_q+
\nonumber\\&& \quad\qquad +\cdots+
          a^\dagger(f_1)a^\dagger(f_2)
          \cdots a^\dagger({\bf h} f_n)\Omega_q
\end{eqnarray}
\end{mathletters}
Then it is elementary to check the following relations:
\begin{mathletters}
\begin{equation}
               \Gamma_q(R^*)=\Gamma_q(R)^*
\end{equation}
\begin{equation}
          \Gamma_q(R)\Gamma_q(S)=\Gamma_q(RS)
\end{equation}
\begin{equation}
  {\rm d}\Gamma_q(\lambda{\bf h}+\mu{\bf k})
    =\lambda{\rm d}\Gamma_q({\bf h})+\mu{\rm d}\Gamma_q({\bf k})
\end{equation}
\begin{equation}
         e^{{\rm d}\Gamma_q({\bf h})}=\Gamma_q(e^{{\bf h}})
\end{equation}
\end{mathletters}
{}From the definition (\ref{e3}) of the operator $S_q$ it is clear
that, for $-1<q<1$,
\begin{eqnarray} \label{e7}
 \Gamma_q(R)&&= S_q^{-1}\Gamma_0(R)S_q
\\  \text{and}\qquad
   {\rm d}\Gamma_q({\bf h})&&= S_q^{-1}{\rm d}\Gamma_0({\bf h})S_q
\quad.\nonumber
\end{eqnarray}
Hence $\Gamma_q(R)$ and $\Gamma_0(R)$ are similar. Using the results
of \cite{Dykema} one can even define a {\it unitary} operator $S_q'$
satisfying equation (\ref{e7}), but we will not need this fact.

\section{Quon Statistical Mechanics}

The density matrix of the grand canonical ensemble of a free Quon
gas with one-particle Hamiltonian ${\bf h}$ is proportional to
\begin{displaymath}
e^{\textstyle -\beta({\rm d}\Gamma_q({\bf h})-\mu N)}
   = \Gamma_q\left(e^{\textstyle
                             -\beta{\bf h}+\beta\mu}\right)
\end{displaymath}
and the grand canonical partition function is the trace of this
expression. Hence the fundamental computation on which the theories
Bose, Fermi, or Quon gases can be built is the evaluation of
${\rm tr}\,\Gamma_q(R)$. In the Bose and Fermi cases the answer is
in every textbook (though maybe not quite in this form):
\begin{mathletters}\label{e8}
\begin{equation} \label{e8a}
      {\rm tr}\,\Gamma_{+1}(R)=\exp{\rm tr}\,\ln{1\over1-R}
\end{equation}
\begin{equation} \label{e8b}
      {\rm tr}\,\Gamma_{-1}(R)=\exp{\rm tr}\,\ln(1+R)\quad.
\end{equation}
\end{mathletters}
With $R=\exp(-\beta{\bf h}+\beta\mu)$ these formulas give back the
standard partition functions for quantum gases. Since the scalar
products (\ref{e2}) depend continuously on $q$ it is natural to
expect for ${\rm tr}\,\Gamma_q(R)$ a formula interpolating between
the two expressions above.

For computing this expression we make use of the operator $S_q$
introduced in equation (\ref{e3}).
Let $P_n$ denote the projection on the $n$-particle space in either
full Fock space or $q$-Fock space, i.e.\ the projection onto the
eigenspace of $\Gamma_q(\openone)$ or $\Gamma_0(\openone)$ with
eigenvalue $n$. Then $S_q P_n=P_n S_q$, and we have
\begin{eqnarray} \label{e9}
       {\rm tr}\,\bigl(P_n\Gamma_q(R)\bigr)
      &&={\rm tr}\,\bigl(P_n S_q^{-1}\Gamma_0(R) S_q\bigr)
\nonumber\\
      &&={\rm tr}\,\bigl(P_n S_q S_q^{-1}\Gamma_0(R)\bigr)
\nonumber\\
      &&={\rm tr}\,\bigl(P_n\Gamma_0(R)\bigr)
\quad.
\end{eqnarray}
This computation shows that ${\rm tr}\,\Gamma_q(R)$ is independent
of $q$. Moreover, $P_n\Gamma_0(R)$ is just the $n$-fold tensor
product of $R$ with itself. Since the trace of a tensor product of
operators is the product of the traces, we obtain
\begin{mathletters}
\begin{equation} \label{e10a}
   {\rm tr}\, P_n\Gamma_q(R)   = \bigl({\rm tr}\, R\bigr)^n
\end{equation}
\begin{equation} \label{e10b}
   {\rm tr}\, \Gamma_q(R)  = {1\over 1-{\rm tr}\, R}     \quad.
\end{equation}
\end{mathletters}
Thus we have found the analog of equations (\ref{e8}). However, in
spite of the continuity of (\ref{e2}) in $q$, the right hand side of
(\ref{e10b}) does not interpolate between (\ref{e8a}) and (\ref{e8b}),
since for $-1<q<1$ it does not even depend on $q$. Consequently, the
grand canonical partition function
\begin{eqnarray} \label{e11}
 {\cal Q}({\bf h},\beta,\mu)
       &&={\rm tr}\, e^{\textstyle -\beta({\rm d}\Gamma_q({\bf h})-\mu N)}
 \nonumber\\
       &&= {\rm tr}\,\Gamma_q\left(e^{\textstyle -\beta{\bf h}+\beta\mu}\right)
\nonumber\\
       &&={1\over \displaystyle 1-e^{\beta\mu}{\rm tr}\, e^{-\beta{\bf h}}}
\end{eqnarray}
and the canonical partition function
\begin{equation} \label{e12}
 {\cal Z}({\bf h},\beta,N)={\rm tr}\, P_N \Gamma_q\bigl(e^{-\beta{\bf h}}\bigr)
       =\left({\rm tr}\, e^{-\beta{\bf h}} \right)^N
\end{equation}
of the free Quon gas do not depend on $q$.

Moreover, the probability distributions of second quantized
one-particle operators ${\rm d}\Gamma_q({\bf k})$ in the
corresponding ensembles are independent of $q$. In the classical
case this would follow from the independence of the partition
functions, since all moments of such probability distributions can
be obtained by differentiating the partition function with respect
to suitable parameters in ${\bf h}$. In Quantum Statistical
Mechanics this works only for the first moment, and fails in general
since ${\rm tr}\, \exp(A+B)\neq{\rm tr}\,\bigl(\exp A \exp B\bigr)$.
For quasi-free states, i.e.\ for the equilibrium states of free
evolutions, however, the conclusion is still valid: the Fourier
transform of the probability distribution of
${\rm d}\Gamma_q({\bf k})$ is
\begin{eqnarray} \label{e13}
 \lambda
    &&\mapsto {{\rm tr}\, e^{i\lambda{\rm d}
                    \Gamma_q({\bf k})}\Gamma_q(R)
               \over {\rm tr}\,\Gamma_q(R)}
\nonumber\\
\nonumber\\
     &&= {\displaystyle {\rm tr}\, \Gamma_q\left(e^{i\lambda{\bf k}}
                                           \right)\Gamma_q(R)
        \displaystyle \over {\rm tr}\,\Gamma_q(R)}
\nonumber\\
    &&= {\displaystyle {\rm tr}\, \Gamma_q\left(
                                         e^{i\lambda{\bf k}}R\right)
        \displaystyle \over {\rm tr}\,\Gamma_q(R)}
\quad,
\end{eqnarray}
and, by (\ref{e10b}), the right hand side of this equation
does not depend on $q$.

There is an algebraic way of looking at these structures.
For $\vert q\vert<1$ it follows from the relations (\ref{e1}) that
$a^\dagger(f)$ is a bounded operator, and we may look at the norm
closed algebra (C*-algebra) generated by these operators in any
particular realization. The C*-algebras are especially important
when one wishes to consider the relations (\ref{e1}) without
necessarily assuming the existence of a vacuum vector $\Omega_q$. It
turns out that the C*-algebra generated by the $a^\dagger(f)$ is
essentially independent of the value of $q$, so that the
insensitivity of Quon Statistical Mechanics to the value of $q$ has
its counterpart in the insensitivity of these algebras. More
precisely, for $\vert q\vert <\sqrt2-1\approx.41$ it was shown in
\cite{QCR} that the C*-algebra generated by the Fock representation
contains the full algebraic information, i.e.\ that every other
realization of the relations (\ref{e1}) either generates an algebra
isomorphic to the Fock representation, or to the quotient of the
Fock representation modulo the compact operators in Fock space.
Moreover, the C*-algebras of the Fock representations for different
$q$ are all isomorphic in the range $\vert q\vert<\sqrt2-1$. For
$q=0$ the algebra obtained in Fock space is known as the
Cuntz-Toeplitz algebra, and its quotient by the compact operators as
the Cuntz algebra \cite{Cuntz}, and is a well-studied mathematical
structure. In the results \cite{QCR} the restriction on $q$ is
expected to be only technical, and they are conjectured to be true
for all $\vert q\vert<1$\ \cite{QCS}. Somewhat sharper results can
be obtained, when one considers only the Fock representations to
begin with \cite{Dykema}.

These algebraic results imply that for $-1<q<1$ the only requirement
we may impose in addition to equation (\ref{e1}) without making the
system contradictory is to demand the strict positivity of
$\sum_ia^\dagger(e_i)a(e_i)$, singling out the Cuntz, rather than
the Cuntz-Toeplitz algebra. This is in stark contrast to the
situation at $q=\pm1$: there the operators in Fock space satisfy the
additional relation
\begin{equation} \label{e14}
  a(f)a(g)-q\,a(g)a(f)=0
\quad,
\end{equation}
which is thus seen to be consistent with (\ref{e1}) for $q=\pm1$.
There are, of course, other realizations of the relations (\ref{e1})
at $q=\pm1$ not satisfying (\ref{e14}). The reason (\ref{e14}) is
satisfied in Boson/Fermion Fock space is the same as the reason
for the difference between Boson/Fermion and Quon Statistical
Mechanics, namely the degeneracy of the $q$-dependent scalar product
(\ref{e2}) at the points $q=\pm1$.

\section{Gibbs' Paradox}

The partition function (\ref{e11}) has a strange property: let us
assume for definiteness that the single Quon is a non-relativistic
spinless particle of mass $m$ in a $d$-dimensional region of volume
$V$. The Hamiltonian is then the Laplacian ${\bf h}=-(1/2m)\Delta$
with Dirichlet boundary conditions. Then, provided the volume is
large enough, and has a decent boundary, Weyl's formula for the
asymptotic behaviour of the eigenvalues of the Laplacian (e.g.\
Theorem XII.78 in \cite{RSimon}) gives (for large volumes:)
\begin{equation} \label{e15}
 {\rm tr}\, e^{-\beta({\bf h}-\mu\openone)}
     \approx e^{\beta\mu} \lambda^{-d} V
\quad,
\end{equation}
where $\lambda=(2\pi\beta\hbar^2/m)^{1/2}$ is the thermal de Broglie
wavelength.
For the expectation value of the particle number we get
\begin{equation} \label{e16}
   \langle N\rangle = {1\over\beta}\,{d\over d\mu}
                       \ln{\cal Q}({\bf h},\beta,\mu)
                    = {V\over e^{-\beta\mu}\lambda^d- V}
\quad.
\end{equation}
Thus $\langle N\rangle$ is not proportional to the volume. Moreover,
at the volume $e^{-\beta\mu}\lambda^d$ the expected particle
number, and indeed the partition function itself diverges. This is
in flat contradiction to the assumptions made in the interpretation
of the grand canonical ensemble as the density matrix of a small
subsystem in interaction with a large system serving as a reservoir
for heat and particles. If there is an upper limit to the size of
Quon gas containers, the idea of a ``particle bath'' breaks down.

It is therefore necessary to go back to the canonical ensemble,
i.e.\ to consider systems with fixed particle number. From
(\ref{e15}) and the canonical partition function (\ref{e12})
we get the Helmholtz free energy
\begin{equation} \label{e17}
 F(\beta,V,N)
    =-{N\over\beta}\ln\left(\lambda^{-d}\,V \right)
\quad,
\end{equation}
but we have not got rid of the paradox, since now the free energy
at fixed density $\rho=N/V$ contains a term growing like $N\ln N$.
Finally, we could go back to the microcanonical ensemble, where we
would find that the entropy fails to be an extensive quantity.

In the Statistical Mechanics of large classical systems this problem
is well known as Gibbs' Paradox \cite{Erwin,Huang}. In that context
it can be resolved by fixing the overall statistical weight of the
$N$-particle phase space by a suitable convention. The factor
$(h^{3N}N!)^{-1}$ used to do this is ultimately to be determined
from the classical limit of Quantum Statistical Mechanics. The
classical limits of the Fermi and Bose gases coincide in this limit.
However, the classical limit of a Quon gas would revive Gibbs' old
problem, for it does not give back the factor $1/N!$. Since
Classical Statistical Mechanics by itself offers no canonical choice
for the normalization of phase space measure, Gibbs' Paradox in the
classical theory can be put aside as a weakness which may be (and
was, in fact) overcome by a more powerful theory. The situation is
more serious for the Quon gas: here the weight is determined
unambiguously by Quantum Statistical Mechanics, and it comes out
wrong.

Whichever way we put it, the Quon gas does not behave like some
``material'' of which one can have a large, homogeneous sample. The
thermodynamic limit, which implicitly or explicitly is the basis of
equilibrium Statistical Mechanics does not make sense for this
system. Things get worse if we consider problems like stability of
matter. It is well-known that a world of only Bosons would collapse
under its electromagnetic interaction. Quons have an even stronger
tendency to cling together, and this is true even for nearly
Fermionic values $q\approx-1$. So perhaps the most powerful test
supporting the standard assumption $q=\pm1$ is the stability of
matter itself.

\section{Discussion}

The second quantization of Quons described above is based on the
postulate that symmetries of the one-particle system must become
symmetries of the many-Quon system. It is therefore insensitive to
modifications of the creation operators by factors which commute
with all these symmetries. Let $\alpha$ be real, and consider
\begin{equation} \label{e18}
  b(f)=a(f) \Gamma_q(q^\alpha\openone)
        =q^\alpha \Gamma_q(q^\alpha\openone) a(f)
\quad,
\end{equation}
with $\Gamma_q(q^\alpha\openone)=\exp\bigl(\alpha N\ln q \bigr)
                         =q^{\alpha N}$.
Then we get
\begin{equation} \label{e19}
 b(f)b^\dagger(g)=q^{2\alpha (N+1)}\, \langle f,g\rangle
                   +q^{2\alpha+1}\, b^\dagger(g) b(f)
\quad.
\end{equation}
Then for $\alpha=-1/4$ (and a single degree of freedom) we obtain
the $q$-oscillator as introduced by Biedenharn \cite{Bieden}, and
independently (with a different parameter $\tilde q=q^{-1/2}$) by
\cite{Farlane}.
On the other hand, we can use the ``Woronowicz normalization'',
named after his version of the one-dimensional relations, which
appears as a subalgebra of quantum ${\rm SU}(2)$ \cite{Worono},
i.e.\ operators $w(f)=(1-q)^{1/2}a(f)$, satisfying
$w(f)w^\dagger(g)=(1-q)\langle f,g\rangle+q\,w^\dagger(g)w(f)$.
In both of these cases the second quantized operators $\Gamma_q(R)$
and ${\rm d}\Gamma_q({\bf h})$ are simply the same as before, and
all our results carry over. The point here is that the second
quantized observables are defined by the second quantization of
symmetries, and not by explicit expressions in the creation
operators. Although there are such expressions, they are not very
useful, since they are of infinite order in $a(f)$ and
$a^\dagger(f)$ \cite{Green,Stanciu}. Indeed, formulas such as
\begin{equation} \label{e20}
H=\sum_{ij}a^\dagger(f_i)\langle f_i,{\bf h} f_j\rangle a(f_j)
\quad,
\end{equation}
familiar from the second quantization of Bosons or Fermions, make
little sense for Quons. For example, for $q=0$ the above operator
$H$ acts only on the ``first'' particle. One could think of a
suitable average over permutations, but this fails, because there is
no natural unitary action of the permutations group on the Quon Fock
space. Clearly, for some Hamiltonians defined by fixed algebraic
expressions in creation and annihilation operators the partition
functions will depend on $q$. We leave open the question whether
there is any sensible second quantization scheme of this type.

By applying the rules of Statistical Mechanics we have restricted
consideration to Quons in equilibrium. Of course, one can avoid our
conclusions if one assumes, firstly, that in the early universe for
some reason there were no Quons, and, secondly, that for all Quons
$q$ is so close to $\pm1$ that the interaction on the cosmic time
scale is not sufficient to bring the Quonic degrees of freedom into
equilibrium. Such arguments have been suggested by Greenberg and
Mohapatra \cite{Greenc}. The cosmological questions raised by this
way out of the dilemma of Gibbs' Paradox are beyond the scope of the
present paper.

Discussing the Gibbs' Paradox  Greenberg \cite{Greena} compares Quon
Statistical Mechanics ($q=0$) to the Statistical Mechanics of
para-Bosons and para-Fermions \cite{Ohnuki}. It is well-known that
para-Bosons of finite order can be considered as Bosons with a
finite number of internal degrees of freedom (and similarly for
para-Fermions). In this analogy the Quons, as particles with
infinite statistics, might be expected to behave like particles with
infinitely many internal degrees of freedom. This would then reduce
Quon Statistical Mechanics to the Statistical Mechanics of ordinary
systems. Greenberg does not make this connection explicit, and it
can be expected that an attempt to do this would lead into
difficulties. The crucial problem is the dynamical behaviour of the
internal degrees of freedom. Even in the para-Boson case this is a
problem, when one assumes that there is absolutely no interaction
between the internal and the external degrees of freedom. The
internal degrees of freedom then never go into equilibrium, and the
equilibrium of the external degrees of freedom depends on the prior
history of the system. For example, when initially all particles are
in the same internal state, there is no difference between this
system and ordinary Bosons. A possible way out is the assumption
that the internal degrees of freedom are always completely
randomized. The partition functions are then the same as for a
one-particle Hamiltonian all of whose eigenvalue multiplicities are
multiplied with the same factor. Clearly, the Statistical Mechanics
of such a system is not qualitatively different from that of the
original system, and does not exhibit Gibbs' Paradox.

However, this way out is impossible for infinitely many internal
degrees of freedom. Infinitely many degrees o freedom cannot be
completely randomized, because the unit operator cannot be
normalized as a density matrix. If one makes an arbitrary choice,
like e.g.\ an equilibrium state for a suitable Hamiltonian for the
internal degrees of freedom, the system will effectively behave like
one with finitely many degrees of freedom, and will not exhibit
Gibbs' Paradox. Consider, for example, Fermions in a spacetime with
additional compactified space dimensions, as suggested in
\cite{Greenc}. If we include a kinetic energy term associated with
the additional dimensions, the internal degrees of freedom have an
equilibrium state due to the compactness of the added dimensions.
This system is well-behaved from the point of view of Statistical
Mechanics, and does not exhibit Gibbs' Paradox. On the other hand,
if we do not include the additional kinetic energy term, there will
be no equilibrium states at all. Thus Quon Statistical Mechanics
appears to be qualitatively different from the Statistical Mechanics
of Bosons or Fermions with (finitely or infinitely many) internal
degrees of freedom.

The Gibbs' Paradox concerns the composition of different Quon
systems. This problem also arises in Quon field theory, where one
has to specify how the observable algebras of two subregions combine
into the observable algebra for the whole region. Certainly, for
discussing Quons in the early universe, this problem of Quon field
theory cannot be avoided. It is one of the fundamental results in
general (algebraic) quantum field theory that (massive) particles
with infinite statistics, such as Quons, cannot occur
\cite{Fredenhagen,Greend}. Since this theory proceeds axiomatically
this objection is even valid for interacting Quon systems, and is
essentially independent of the way the Quon observable algebras are
defined in terms of creation and annihilation operators. As an
illustration consider a relativistic field theory of Quons, set up
by taking as one-particle space an irreducible representation space
of the Poincar\`e group, and applying the second quantization scheme
introduced above. If $m\geq0$ for the one-particle representation
the second quantized system will also obey the spectral (positive
energy) condition. In the Fermi-or Bose case the local algebra
associated with a bounded open region in space-time is defined in
terms of the real-linear subspace of the one-particle space,
consisting of the Fourier transforms of the real test functions
supported in the given region \cite{Araki}. The Boson local
field algebra is then generated by the operators $a(f)+a^\dagger(f)$
with $f$ in the associated subspace. It is easy to see from the
relations (\ref{e1}) that the corresponding prescription fails to
produce algebras commuting at spacelike separation. Even restricting
to the gauge invariant subalgebra of the field algebra (as is done
in the Fermi case) will not make these algebras commute. On the
other hand, for commuting operators $R,S$, the operators
$\Gamma_q(R)$ and $\Gamma_q(S)$ also commute. Therefore, if we could
associate with each space-time region a complex-linear subspace such
that spacelike regions would correspond to orthogonal subspaces, we
could generate commuting local algebras by the operators
$\Gamma_q(R)$ with $R$ supported by the appropriate subspace. It is
well-known that such a net of orthogonal subspaces can be found, as
soon as one replaces the irreducible representation by a reducible
one, containing both negative and positive frequencies. In that
case, however, the field theory would no longer obey the spectral
condition.

\section*{Acknowledgement}
This paper grew out of a collaboration \cite{QCR} with Lothar
Schmitt (Osnabr\"uck) and Palle J\o rgensen (Iowa). It has benefited
from conversations with R.~Speicher, B.~K\"ummerer, A.~Nica,
K.~Dykema, H.~Maassen, and K.-H.~Rehren. The author acknowledges
financial support from the DFG (Bonn) through a scholarship and a
travel grant.

\end{document}